\documentstyle[11pt,paspconf,epsf]{article}

\begin{document}

\title{Optical Trapping by Radiometric Flow}
\author{William L. Clarke}
\affil{Department of Physics, University of California at Santa Cruz wlclarke@cats.ucsc.edu}

\begin{abstract}
Micron sized, neutral, non-dielectric particles immersed in a viscous fluid
can be trapped in the focal plane of a Gaussian beam.
A particle can absorb energy from such  a beam with a large radial intensity
gradient, resulting in substantial temperature gradients and a radiometric torque
 which causes it to spin rapidly about an axis perpendicular to the flux of radiant
energy. The particles are also observed to orbit about the optical axis.
Here we investigate the fundamental physics of this system, the Radiometric 
Particle Trap, and discuss its force laws using gas-kinetic theory.
\end{abstract}


\section{Phenomenology}

Figure 1.  is a high resolution optical micrograph of a submicron sized, spinning
particle orbiting around the optical axis of a focused laser beam. In fact, the
particle is trapped in a stable orbit in the focal plane of the beam by an
interaction between the particle's spin moment and the large radial optical 
intensity gradient characteristic of a focused, Gaussian coherent light source.

\begin{figure}[h]
\epsfxsize=8.5cm
\epsfysize=7cm
\centerline{\epsffile{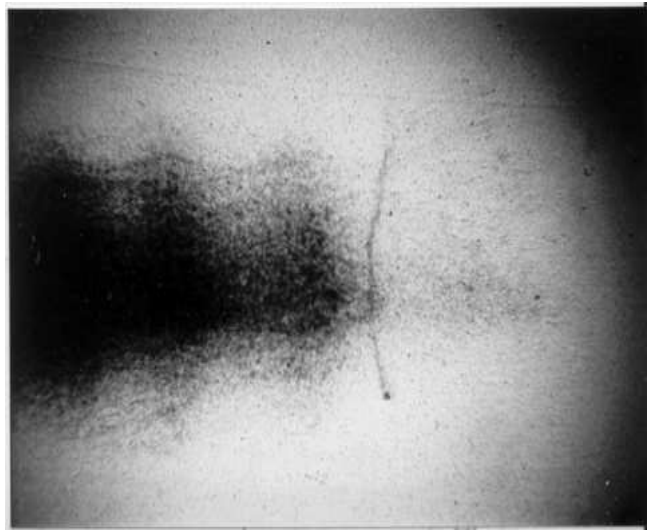}}
\caption[1]{\label{f1} Graphite particle in orbit about the optical axis, immersed in 1/2 atm. nitrogen.  (1/500 sec at 1000x). }
\end{figure}

This phenomenon was first discovered by Steve Wilson [1] in 1965, when he
trapped micron sized, non-dielectric particles in the focal plane of a focused
Gaussian laser beam. The techniques used are described in detail in his 
monograph on experimental microscopy [2].

The particles are contained in a glass cylinder at the focus of a Gaussian beam, and immersed in nitrogen gas at approximately 1/2 atm. The particles are observed to drop out of the beam when the cylinder is pumped down to approximately 0.1 atm. Hence, the mechanism of trapping is distinct from the mechanism of Ashkin [3], and Chu
 [4], involving a gradient force on optically transparent, dielectric particles, which obtains in a vacuum. 
Rather, the photograph above shows a spinning graphite particle which is 
trapped by \emph{radiometric flow} [5], a fluid dynamical regime which depends on the
large temperature gradients induced on the particle by the absorption of optical
 energy from the beam. The surface temperature gradients induced on the particle
 by the beam cause radiometric forces which drive the system. We note that the 
radiometric forces are completely due to the non-equilibrium condition of the hot gas around the particle.

The next photograph shows three spinning, graphite particles. The direction of 
the beam is left to right;  the circular orbit of each particle
lies in a plane normal to the optical axis. It is seen that the spin axis of
each particle is approximately along the direction of its orbit. The particles 
are heated unevenly,
since the part of a particle closer to the optical axis will absorb more
radiant energy than the part further from the axis, and also the side of the particle
closest to the source of the beam (ie the front of the particle) will absorb more
energy than the back. This uneven absorption of optical energy induces a
radiometric moment on each particle which causes it to spin.

\begin{figure}[h]
\epsfxsize=8.5cm
\epsfysize=7cm
\centerline{\epsffile{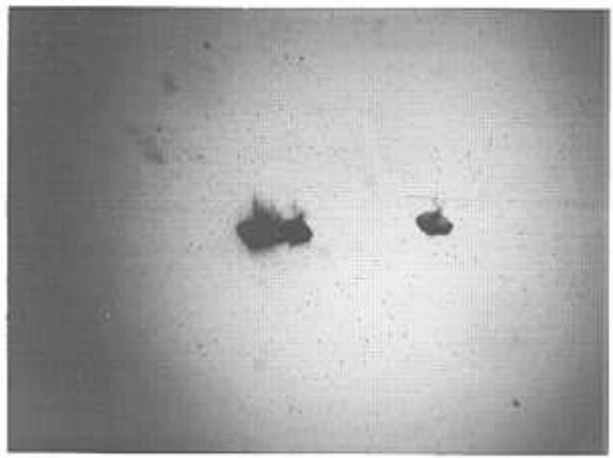}}
\caption[1]{\label{f2} Three distinct carbon particles spinning on axes approximately
parallel to their orbital motion. The optical axis is from left to right, and
 each particle is orbiting in a plane normal to the optical axis. (1/1500 sec at 2200x) }

\end{figure}

Furthermore, additional temperature gradients are induced along the particle spin
axis which causes it to orbit. The spinning particles exhibit a \emph{dynamic chirality} \ k(t) \ which couples the spin degree of freedom to the orbital motion.  We have deduced the radiometric force laws which
cause the particles to spin and orbit from the stready solution of radiometric flow
for an ellipsoid (See Section 2). Indeed, a well-defined, small dimensional dynamical 
system may be derived from these force laws which can be put on the computer, and 
integrated. We find that the simulated system  admits limit cycle solutions which are
periodic and stable. The existence of limit cycle solutions in the simulations
verifies 
the stability criterion which was derived from pure analysis [6]. We conclude that the
existence of spinning, orbiting particles trapped in a Gaussian beam is a
nonlinear mode of a radiometric-mechanical system of particles immersed in a viscous
fluid.         

Figure 3. shows a simulation of a 1 micron particle, initially at a 
distance of 8 microns from the optical axis of a focused Gaussian beam with spot 
diameter 5 microns. The particle has zero initial spin angular velocity $\omega_0$, and zero
initial orbital angular velocity $\Omega_0$. The large intensity gradient of the beam
immediately induces a large radiometric moment on the particle, causing it to spin,
and orbit the optical axis.

\begin{figure}[h]
\caption[1]{\label{f2} is attached, due to formatting difficulties} 
\end{figure}

The particle spirals into a limit cycle [7]. In fact any particle within approximately
3 times the beam spot radius will spiral into the same steady, stable orbit.
That is, the limit cycle attractor is the asymptotic orbit for all such particles,
for any reasonable initial spin and orbital angular velocities.

The next photograph (Figure 4) shows multiple clouds of graphite particles which  
are trapped near the focal plane. It is observed that the clouds repel each other, yet
yet appear to be electrostatically neutral. This can be seen by the observed null effect
of electrostatically charged probes brought into the vicinity of the orbiting
 clouds.

\vspace{1.0cm}

(Figure 4 appears on the next page)

\begin{figure}[h]
\epsfxsize=8.5cm
\epsfysize=7cm
\centerline{\epsffile{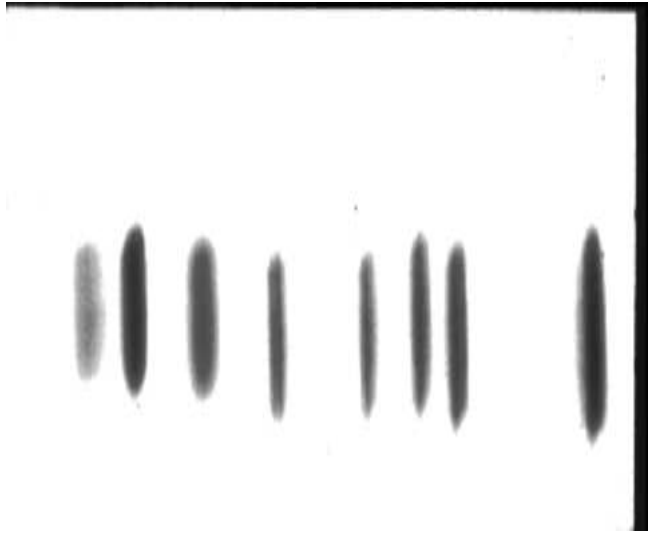}}
\caption[1]{\label{f4} High resolution image of carbon particles immersed in 0.5 atm
 of argon. Orbiting particles form multiple mutually repulsive discrete groups. The beam spot radius \ $\sigma$ $\approx$ 5 microns.

}
\end{figure}

\vspace{1.0cm}

We find that the clouds of particles are pushed towards the focal plane by 
longitudinal forces which act along the optical axis. These are also radiometric forces
which arise from the fact that the surfaces of constant optical intensity of
a Gaussian beam are in fact hyperboloids [8], so that the spin axis of the orbit
ing particles has a longitudinal component. The temperature gradient along this
spin axis drives the particles towards the focal plane, where they form a linear
 array of stable, mutually repulsive clouds.

We therefore claim that the dynamics of multiple clouds of spinning, orbiting 
particles can be accurately described by a system that assumes a non-local interaction
between the clouds. We note that for a 100 mW beam and 5 micron spot radius,
the particles will spin at approximately 100,000 rad/s, and the particle will
orbit with an orbital angular velocity of approximately 3000 rad/s. The motion of
the "bare" particle will then induce a toroidal vortex ring in the fluid around
the orbiting particle. It is known that such vortex rings repel [9] by the laws
of fluid potential flow. Thus we explain the existence of multiple trapped clouds
of particles by the laws of vortex motion. We have derived an N-body force 
law which describes the interaction of N toroidal clouds of particles, trapped 
along the optical axis near the focal plane [10].

\section{Theory}

The theory of this complex physical system must depend on the basic laws of 
radiometric flow. We follow Maxwell [11], who in 1880 derived the correct equations 
of motion and boundary conditions for fluid flow around objects with large 
temperature gradients. We have essentially followed his derivation using more modern 
notation, and we find that we can give a rigorous definition of the regime of
radiometric flow at low Reynolds number,
where the Prandtl number $P_r$ = 1, the particle size is of order the mean free path 
\ $\lambda$ \ of the immersing gas, which is moreover considered to be incompressible, with no
sources or sinks of heat except at the boundaries.

We find that that the flow is identical to Navier-Stokes flow with the non-standard (radiometric)
boundary condition:

\vspace{0.5cm}

(1)\ \ \ \ \ $\vec{v}$  = $\vec{v}_{slip}(\nabla T)$ at the boundaries \ \ \ \ \ where

(2)\ \ \ \ \ $\vec{v}_{slip}(\nabla T)$ \ is a linear function of \ $\nabla T$ \ at the boundaries. 

\vspace{0.5cm}

Maxwell's derivation of this regime depends on the Chapman-Enskog [12] approximation
of the velocity distribution of the gas just outside the particle, which is
in fact non-Maxwellian, ie in a state of non-equilibrium. This non-equilibrium 
distribution is parameterized by 20 expansion coefficients, which may be expressed
in measurable quantities, such as the gas density, temperature, etc. Solutions
to the equations of motion with the correct radiometric boundary conditions have
allowed us to calculate the stress tensor of the immersing gas at the particle
surface, and therefore the radiometric forces and moments on a sphere, ellipsoid
of revolution, and also for a circular flat plate, which is a degenerate case
of the ellipsoid [13].

For example, we may calculate the radiometric flow outside a circular plate normal to the z-axis, with the pure quadrupole temperature distribution

\vspace{0.5cm}

(3)\ \ \ \ \ T  =  T$(v,\phi)$  =  $T_0$  +  ($\delta T_Q$)[ 1/2  +  $G_2(v,\phi))$] \ \ \ \ \ where

(4)\ \ \ \ \ $G_2(v, \phi)$  =  (1/8) (3 $cos^2(v)$ - 1)    
+ (3/4) $sin(2v)$ cos($\phi$)

\ \ \ \ \ \ \ \ \ \ \ \ \ \ \ \ \ \ \ \   + (3/8) $sin^2(v)$ cos(2$\phi$)

\vspace{0.5cm}

This is the leading term of the surface temperature variation of an ellipsoidal particle irradiated by a beam with a radial intensity gradient $\nabla$I. The induced quadrupole moment \ $\delta T_Q$ \ is given by 

\vspace{0.5cm}

(5)\ \ \ \ \ $\delta T_Q$ = (1/4) $a^2$ $\nabla$I / $k_p$ 

\vspace{0.5cm}

where \ $k_p$ \ is the thermal conductivity of the particle.

The derivation of the exact velocity field around the plate requires the use of ellipsoidal coordinates, and is long and tedious. However, we are able to use the solution to calculate the stress tensor in these coordinates, and therefore the net radiometric moment on a circular plate of radius a 

\vspace{0.5cm}

(6)\ \ \ \ \ $\vec{M}$ = $\hat{y}$ \ [ 9$\pi$ $\rho \nu^2$ a $\gamma^{\prime}$ ($\delta T_Q$/T) ]

\vspace{0.5cm}

where \ $\gamma^{\prime}$ \ is a geometrical factor of order unity, and where we have assumed that the optical axis is in the $\hat{z}$ direction. The kinematic viscosity \ $\nu$ = $\eta$ / $\rho$. We find that the spin axis of the particle is perpendicular to the flow of radiant energy of the beam. This radiometric moment then causes the particle to spin with an angular velocity \ $\omega$ $\propto$ $\gamma^{\prime}$ ($\delta T_Q$/T).

Similarly, we may calculate the radiometric force on a circular plate with a dipole temperature variation. The result is

\vspace{0.5cm}

(7)\ \ \ \ \ $\vec{F}$ = $\hat{\phi}$ \ k(t) \ [ 6$\pi$ $\rho \nu^2$ $\gamma$ ($\delta T_D$/T) ]  

\vspace{0.5cm}

The induced dipole moment \ $\delta T_D$ \ is proportional to \ $\nabla$I, and there is a small temperature differential along the spin axis which is proportional to k(t) $\delta T_D$ \ in the $\hat{\phi}$ direction. This is the orbital radiometric force which drives the particle against the orbital viscous drag. The dynamical chirality factor k(t) arises from the angular inertia of the spinning particle, and its law of transformation under rotations [14].

With these solutions, which are exact dipole and quadrupole solutions of the 
equations of motion for steady flow at low Reynolds number, we are able to build a
analytical model of the radiometric particle trap dynamical system, ie we use the
exact expressions for the steady radiometric forces and moments as the force 
laws which are incorporated into a low dimensional mechanical system which captures
the essential dynamics of orbiting, spinning particles trapped by the beam.

The main result of the analysis is the derivation of the so-called \emph{spin-gradient}
central force which holds the particle in its orbit [15]. This force arises as a
 coupling of the particle spin angular velocity to the motion of its c.m. due to
 the temperature  variation of the kinematic viscosity \ $\nu$ $\propto$ $T^{3/2
}$. We find

\vspace{0.5cm}

(8)\ \ \ \ \ $\vec{F}_{s.g.}$ = - $\hat{r}$ [ 18 $\rho \nu^2$
$\gamma^{\prime}$ ($\delta T_Q$/T) $\gamma$ ($\delta T_D$/T) ] $\propto$ $\omega$ $\nabla$I

\vspace{0.5cm}

The typical magnitude of the central acceleration is approximately 50 $m/s^2$ 
$\approx$ 5 g, sufficient to support the particles in a gravitational field. The
combination of the radiometric moment, orbital force, and spin-gradient trapping
force results in steady, stable, and approximately circular orbits.

The derivation of these force laws can be considered to be semi-rigorous,
consistent with the underlying gas-kinetic theory, and dimensionally consistent
with the regime of radiometric flow. We find that we can pin down all the geometrical
coefficients and other factors of order unity, resulting in a dynamical system
which depends on no free parameters, ie every factor of order unity is accounted
for.

\section{Conclusions}

We have discussed the existence of a steady, stable periodic mode of motion of the system of micron sized particles immersed in a viscous fluid, trapped in the focal plane of a Gaussian beam. The fundamental force laws may be derived from first principles, ie gas-kinetic theory and offer an explanation of  the observed phenomena. This is one of the few methods of trapping non-dielectric particles (such as metallic contaminants) and may find important applications in the field of ultra-clean gas flows.   

The system is driven by radiometric forces which arise from the non-Maxwel-
lian distribution of the gas molecules near the surface of the immersed particles caused by the large temperature gradients induced by the beam. Because the Gaussian beam profile falls off so abruptly, the particle surface temperature variation contains a large quadrupole component which causes it to spin. The angular inertia of the spinning particle then results in a small coupling of the particle spin angular momentum into the orbital direction, which sustains its motion against the orbital viscous drag. Finally, the effect of the radial intensity gradient coupled with the particle spin produces an asymmetry of forces in the - $\hat{r}$ direction which causes the spin-gradient central force. We have given quantitative estimates for these forces and moments.

Furthermore, the particle trap theory makes qualitative as well as quantitative
predictions. It is non-obvious why the particles are caused to orbit, since a 
Gaussian beam has rotational symmetry around the optical axis. The simulations show
that any particle with an infinitesimally small intrinsic chirality $k_0$ will result
in the spin of the particle being coupled into an orbital motion. We conjecture
that even a small helical component of radiation pressure arising from an 
infinitesimally small admixture of a Laguerre-Gaussian (helical) mode [16] would 
provide such an infinitesimal intrinsic chirality. Special holograms are available
to generate such modes [17], so that a small intrinsic chirality could be generated
with either sense, which should result in clockwise or counter-clockwise orbits.
The system thus exhibits "dynamical symmetry breaking" of the rotational 
symmetry of the Gaussian beam.

\end{document}